\documentclass[conference]{IEEEtran}

\usepackage{amsmath,amssymb,amsthm}
\usepackage{cite}
\usepackage{array}
\usepackage{graphicx}
\usepackage{url}
\usepackage{booktabs}
\usepackage{xcolor}

\newtheorem{remark}{Remark}

\title{Cryptanalysis of a Lightweight RFID Authentication Protocol Based on a Variable Matrix Encryption Algorithm}

\author{
\IEEEauthorblockN{Hongjun Wu}
\IEEEauthorblockA{Nanyang Technological University, Singapore\\
Email: wuhj@ntu.edu.sg}
}

\begin{document}
\maketitle

\begin{abstract}
Recently, a two-way RFID authentication protocol based on the so-called AM-SUEO-DBLTKM variable matrix encryption algorithm was proposed for low-cost mobile RFID systems. Its design combines adaptive modulus selection, self-updating matrix ordering, and transpose/block-based matrix generation. In this paper, we show that the protocol has fundamental structural weaknesses. First, the underlying encryption primitive remains a linear transformation modulo a session modulus, with no nonlinear confusion layer and no ciphertext chaining. Second, in the concrete lightweight setting emphasized by the original paper, the effective update space is very small: there are only a few modulus choices, only four matrix-order choices when two secret matrices are used, and only a limited family of DBLTKM-generated matrices. Third, the correctness requirements of the protocol impose nontrivial constraints on the sizes of the modulus and plaintext coordinates, which substantially weaken the claimed entropy of the secret quantities.

Building on these observations, we describe a multi-session algebraic attack path. Under repeated reuse of the same effective matrix and modulus---an event that is plausible because of the small update space---ciphertexts corresponding to $N_t$, $N_t+1$, $N_r$, and $N_r+1$ reveal a full column of the effective matrix. Across sessions, transpose-based matrix generation helps recover additional entries of the underlying secret matrices, while the remaining entries can be obtained later from ordinary ciphertext equations. We then show that candidate factors of the session moduli can be tested by solving reduced equations for the long-term secret $S$ across many sessions and checking for repeated consistent solutions. This, in turn, enables recovery of candidate 64-bit moduli and the remaining protocol secrets. Taken together, our results indicate that the protocol is structurally insecure and admits a realistic route to full compromise in the lightweight parameter regime advocated for deployment.
\end{abstract}


\section{Introduction}
RFID authentication for low-cost tags remains difficult because the security requirements and the implementation constraints point in opposite directions. The protocol analyzed in this paper was recently proposed by Wang et al. \cite{target}. On the one hand, an RFID protocol deployed in open wireless environments should resist replay, impersonation, traceability, desynchronization, and active manipulation. On the other hand, passive and low-cost tags have very limited logic area, storage, and online computational capability. This tension has led to a substantial body of work on lightweight and ultra-lightweight RFID authentication protocols \cite{juels2003,weis2004,juels2006,survey2023,scott2024}.

A standard observation in this area is that there are two very different design philosophies. One line of work adapts conventional cryptographic mechanisms to RFID and IoT settings. Examples include protocols based on hash functions, block ciphers, stream ciphers, elliptic-curve cryptography, or other more mature primitives \cite{juels2006,ning2010,cao2018,amin2016,present,skinny,quark}. Such schemes usually inherit better understood security foundations, but they may be criticized for implementation cost when the target is an extremely constrained tag. A second line of work aims at extremely lightweight deployment and therefore replaces mature primitives with simple bitwise operations, table updates, permutations, or algebraically structured transformations \cite{dimitriou2005,perislopez2006,perislopez2006m2ap,perislopez2007emap,cho2007,chien2007,lee2008,kumar2020}. The literature has repeatedly shown that the second direction is fragile: once the primitive becomes too structured, security claims often fail under direct cryptanalysis even if the protocol looks attractive from a cost perspective \cite{juels2006,survey2023,scott2024}.

This issue is particularly visible in ultra-lightweight RFID protocols. Early proposals such as LMAP, M2AP, and EMAP were motivated by the need to avoid expensive cryptographic components on low-cost tags \cite{perislopez2006,perislopez2006m2ap,perislopez2007emap}. Similar motivations appear in later authentication designs for IoT, medical, and logistics environments \cite{ning2010,cao2018,amin2016,lbraps2022}. However, the history of the field also shows that reducing a protocol to a small number of linear or near-linear operations often introduces strong algebraic relations that are exploitable. In that sense, storage efficiency and gate-count efficiency do not automatically imply security; in some cases they are achieved precisely by introducing the regularity that makes the system vulnerable.

Besides protocol-level lightweight design, another research direction uses hardware characteristics such as physically unclonable functions. PUF-based schemes attempt to strengthen authentication without storing explicit long-term secrets in ordinary memory \cite{bolotnyy2007,devadas2008,puf2017,puf2021}. These approaches are important, but they rely on a different threat model and different implementation assumptions. In many practical settings, especially for very cheap tags, protocol designers still seek software-like or arithmetic-only constructions. That is exactly the context in which structured matrix-based designs continue to appear.

The protocol studied in this paper is a recent example of this latter design style. It is built around modular matrix multiplication rather than around a standard cryptographic primitive. The protocol combines three mechanisms: adaptive modulus selection (AM), self-updating encryption order (SUEO), and diagonal block local transpose key matrix generation (DBLTKM). The intended effect is clear. Starting from a very small number of secret matrices, the protocol attempts to create many apparent encryption states by varying the modulus, varying the order of multiplication, and allowing transpose- or block-based matrix generation. This idea is attractive only if those generated states are both sufficiently numerous and sufficiently independent from a cryptanalytic point of view.

Our view is that this is exactly where the design fails. The first reason is conceptual. The core encryption rule is still only a linear map modulo a selected modulus. The second reason is combinational. In the lightweight setting emphasized by the original protocol, the number of secret matrices must remain small, and therefore the number of genuinely different effective states is also small. The third reason is mathematical. The protocol description mixes 128-bit secrets and nonces with 64-bit matrix elements and modular decryption, which creates nontrivial correctness constraints that are not addressed carefully in the original design. Once these constraints are taken into account, the claimed entropy of the system becomes much smaller than it first appears.

This problem is not isolated to RFID. There is a long history of cryptanalysis of matrix-based encryption, beginning with the Hill cipher and continuing through later variants that attempted to strengthen it by modifying the key schedule, the modulus, or the plaintext representation \cite{hill1929,overbey2005,cryptanalysis_hill2,cryptanalysis_hill1,toorani2008,islam2012hill}. The broad lesson from this literature is well known: if the encryption operation remains essentially linear, then apparent key-space growth obtained by rearranging or composing linear maps often does not translate into real security. This historical perspective is directly relevant here because the AM-SUEO-DBLTKM design ultimately derives all of its states from a very small family of structured matrices.

Another point worth emphasizing is methodological. RFID papers often include symbolic or logic-based validation, for example BAN-logic arguments or automated protocol checks, to support claims of mutual authentication and resistance to replay \cite{banlogic,avispa2005}. Such analyses can be useful for checking message-flow assumptions under idealized primitives. They do not, however, establish the hardness of attacking the underlying arithmetic transformation. If the encryption primitive itself is weak, then symbolic validation does not save the protocol.

In this paper, we analyze the AM-SUEO-DBLTKM-based RFID protocol from the perspective of direct cryptanalysis of the underlying matrix mechanism. Our starting point is the concrete lightweight regime highlighted in the original work, namely the case where only two secret matrices are used. In that regime, the SUEO mechanism yields only four orderings, the DBLTKM mechanism still produces only a limited family of effective matrices, and the update values contribute only a few bits of effective control entropy because they are used only through modular reductions into very small tables. We also show that the correctness of decryption imposes a much tighter range on the plaintext coordinates than the nominal 128-bit statement suggests.

After identifying these structural weaknesses, we present a multi-session attack. Very roughly, if the same effective matrix and modulus are reused to encrypt related values such as $N_t$, $N_t+1$, $N_r$, and $N_r+1$, then ciphertext differences reveal an entire column of the effective matrix. Since DBLTKM explicitly allows transposed variants, repeated observations across sessions can be combined to recover three entries of each underlying secret matrix, after which the remaining entry can be obtained from an ordinary ciphertext equation. We then test candidate reduced modulus values by solving for the long-term secret $S$ across many sessions. Candidates that are compatible with the true modular structure produce repeated and mutually consistent values of $S \bmod q_x$, whereas incompatible candidates do not show such stable behavior. This narrows the set of plausible session moduli and, after full verification, leads to recovery of the remaining protocol state.

The main contribution of this work is therefore not merely another attack on a specific RFID proposal. Rather, it is a case study showing that in extremely lightweight authentication design, combinational growth of matrix states should not be confused with cryptographic strength. When the underlying primitive is linear and the number of truly independent secret objects is very small, multi-session algebraic attacks become natural.

\subsection{Our Contributions}
We summarize our contributions as follows.
\begin{enumerate}
    \item We revisit the parameter regime emphasized for lightweight deployment and show that the concrete construction is necessarily very small. In particular, the protocol effectively relies on two secret matrices in the lightweight setting, leading to only four possible matrix-order choices and only a limited family of DBLTKM-generated matrix states. 
    \item We identify a correctness inconsistency in the original description. Since encryption and decryption are performed modulo a 64-bit modulus while nonces, identities, and secrets are described as 128-bit values, exact recovery is impossible unless the plaintext coordinates are constrained to fit below the modulus. This effectively reduces the recoverable entropy of each 128-bit quantity.
    \item We give a multi-session algebraic attack on the effective matrices. When the same effective matrix and modulus are reused to encrypt $N_t$, $N_t+1$, $N_r$, and $N_r+1$, ciphertext differences reveal a full column of the effective matrix. Repeating the observation across sessions and exploiting transpose-based DBLTKM updates yields recovery of three entries of each underlying secret matrix, while the remaining entry is recovered later from an ordinary ciphertext equation.
    \item We propose a modulus-recovery phase based on multi-session consistency tests. By testing candidate reduced modulus values $q_x$ and solving for the long-term secret $S$ over many sessions, the attacker can identify candidates that repeatedly produce consistent reduced solutions. These candidates can then be combined into candidate 64-bit session moduli and verified against the full equations, leading to recovery of the remaining protocol secrets.
\end{enumerate}

\subsection{Paper Organization}
The rest of the paper is organized as follows. Section~\ref{sec:overview} summarizes the original protocol and highlights the parameter choices relevant to our attack. Section~\ref{sec:prelim} discusses the effective security space and the correctness constraints implied by modular decryption. Section~\ref{sec:matrix-recovery} presents the matrix-recovery phase. Section~\ref{sec:modulus} describes the recovery of factors of the session moduli and the reconstruction of the candidate $q$ values. Section~\ref{sec:fullbreak} explains how the remaining secrets can be recovered, leading to a complete break. Section~\ref{sec:conclusion} concludes the paper.

\section{Overview of the Target Protocol}\label{sec:overview}
The target protocol is built on the linear encryption primitive
\begin{equation}
E(t,A,p)=A t \bmod p,
\label{eq:basicenc}
\end{equation}
and the corresponding decryption rule using a modular inverse matrix. The protocol paper states that nonces, secret values, modulus, and identifiers are 128-bit values, while each matrix element is 64 bits. The protocol uses two secret matrices in its concrete lightweight example and derives effective encryption states by three mechanisms:
\begin{itemize}
    \item \textbf{AM:} choose a modulus $q$ from an AM index table;
    \item \textbf{SUEO:} choose an ordering of the secret matrices;
    \item \textbf{DBLTKM:} generate transpose/block-based variants of the matrices.
\end{itemize}

In the authentication stage, the protocol transmits, among others, the encryptions of $N_t\|S$, $N_r\|S$, $N_r\|S^d\|S^p\|S^c$, $N_t\|S^d\|S^p\|S^c$, $N_t+1\|ID$, and $N_r+1\|ID$. The long-term secret $S$ is reused across sessions, while $N_t$ and $N_r$ are fresh random numbers generated by the tag and reader, respectively. The server further generates update values $S^d,S^p,S^c$, which are subsequently reduced modulo small table sizes to select the next DBLTKM pattern, the next SUEO order, and the next AM modulus.

\subsection{Protocol Flow Relevant to the Attack}
For completeness, we summarize only the parts of the target protocol that are directly relevant to our cryptanalysis. Let $A$ denote the current effective encryption matrix and let $p$ denote the current modulus in the pre-update phase. After the server generates the update values $S^d$, $S^p$, and $S^c$, the protocol derives a new effective matrix $A_{\mathrm{new}}$ and a new modulus $q$ through the DBLTKM, SUEO, and AM mechanisms. The corresponding decryption matrix is denoted by $B_{\mathrm{new}}$.

The protocol messages used in our attack can be summarized as follows.

\begin{enumerate}
    \item The tag generates a fresh random nonce $N_t$ and sends
    \begin{equation}
    C_1 = E(N_t\|S, A, p).
    \end{equation}

    \item The reader generates a fresh random nonce $N_r$ and sends to the server
    \begin{equation}
    C_2 = E(N_r\|S, A, p).
    \end{equation}

    \item After authenticating the reader, the server generates new secret values $S^d$, $S^p$, and $S^c$, and sends
    \begin{equation}
    C_3 = E(N_r\|S^d\|S^p\|S^c, A, p).
    \end{equation}

    \item The reader forwards to the tag
    \begin{equation}
    C_4 = E(N_t\|S^d\|S^p\|S^c, A, p).
    \end{equation}

    \item Using the update values, the parties compute
    \begin{equation}
    \begin{aligned}
    S^d \bmod Z_{\mathrm{DBLTKM}},\quad
    S^p \bmod Z_{\mathrm{SUEO}},\\
    S^c \bmod Z_{\mathrm{AM}}.
    \end{aligned}
    \end{equation}
    and thereby derive the next effective encryption state $(A_{\mathrm{new}},q)$.

    \item The tag then sends
    \begin{equation}
    C_5 = E(N_t+1\|ID, A_{\mathrm{new}}, q),
    \end{equation}
    and the reader forwards to the server
    \begin{equation}
    C_6 = E(N_r+1\|ID, A_{\mathrm{new}}, q).
    \end{equation}
\end{enumerate}

The attack developed in this paper uses two kinds of structure from the above flow. First, the messages $C_1$, $C_2$, $C_3$, and $C_4$ are all encrypted under the same pre-update state $(A,p)$. Second, the messages $C_5$ and $C_6$ are encrypted under the same post-update state $(A_{\mathrm{new}},q)$. This separation allows us to analyze repeated effective matrices and moduli across sessions and to exploit the algebraic relations between encryptions of $N_t$, $N_t+1$, $N_r$, and $N_r+1$.

\subsection{Small Concrete State Space}
The lightweight example emphasized by the protocol paper takes the number of key matrices to be $N=2$. For this setting, the protocol itself yields the following state-space sizes:
\begin{align}
Z_{\mathrm{SUEO}} &= \binom{2}{1}1!+\binom{2}{2}2! = 4,\\
Z_{\mathrm{DBLTKM}} &= (2N)+(2N)^2 = 4+16 = 20.
\end{align}
Moreover, the accompanying example uses an AM table with eight modulus choices. Hence the entire update mechanism is built from a very small number of combinations. This fact is central to our attack.

For $N=2$, the SUEO choices are precisely
\begin{equation}
A,\quad B,\quad AB,\quad BA.
\end{equation}
Thus the first matrix applied to a plaintext block is either $A$ or $B$, with probability roughly $1/2$ under a uniform model. The uncertainty introduced by SUEO is therefore extremely limited in the lightweight regime.

\section{Effective Security Space and Correctness Constraints}\label{sec:prelim}
In this section we show that the protocol's effective security space is much smaller than the raw bitlengths in the description suggest.

\subsection{Tiny Control Entropy of the Update Secrets}
Although the protocol describes $S^d$, $S^p$, and $S^c$ as 128-bit values, they are used only through
\begin{equation}
S^d \bmod Z_{\mathrm{DBLTKM}},\qquad
S^p \bmod Z_{\mathrm{SUEO}},\qquad
S^c \bmod Z_{\mathrm{AM}}.
\end{equation}
For the concrete lightweight setting, this means that the update secrets contribute only
\begin{equation}
\log_2 20 \approx 4.32,\qquad \log_2 4 = 2,\qquad \log_2 8 = 3
\end{equation}
bits of control entropy, respectively. Therefore, the nominal 128-bit sizes of $S^d,S^p,S^c$ do not translate into comparable cryptographic strength.

\subsection{DBLTKM-Generated Matrix Family}
For the concrete setting $N=2$, the protocol paper gives
\begin{equation}
Z_{\mathrm{DBLTKM}}=20,
\end{equation}
which should be understood as the size of the DBLTKM construction space in the illustrative lightweight example. Together with the AM and SUEO mechanisms, this still yields only a small family of effective matrices for the protocol. This limited state space is important for cryptanalysis, because repeated effective states across sessions are far more plausible than they would be in a conventional design with a large independent key space.

\subsection{Correctness Imposes a Reduced Plaintext Range}
The protocol description states that nonces, secrets, identities, and the modulus are 128-bit values, while each matrix element is 64 bits and encryption/decryption are performed modulo the selected modulus. However, a plaintext coordinate can be recovered exactly from modular decryption only if it is represented in a range strictly below the modulus. If an arbitrary 128-bit value is encrypted directly modulo a 64-bit modulus, decryption can recover only the residue class, not the original integer.

Therefore, correctness requires an implicit encoding of each 128-bit quantity into smaller coordinates. In the most natural two-coordinate encoding, each coordinate must be smaller than $q$. If $q$ is chosen as a full 64-bit modulus near $2^{64}$, then a safe correctness margin requires each coordinate to be at most about 63 bits, meaning that the recoverable entropy per nominal 128-bit value is effectively about $63\times 2=126$ bits rather than an unrestricted 128 bits. Likewise, the effective matrix entries must be interpreted as values constrained so that modular decryption remains meaningful.

This observation has two cryptanalytic consequences. First, it weakens the entropy claims made by the target protocol. Second, it changes the behavior of recovered matrix columns: when the recovered column entries are below the session modulus, they may appear without visible wraparound, which means that modulus recovery must rely on consistency across sessions rather than single-session residue anomalies.

\section{Recovery of the Effective and Secret Matrices}\label{sec:matrix-recovery}
We now present the first phase of the attack. The starting point is the linearity of the encryption primitive.

\subsection{A Column-Recovery Observation}
Consider two encryptions under the same effective matrix $M$ and the same modulus $q$:
\begin{equation}
C_1 = M T_1 \bmod q, \qquad C_2 = M T_2 \bmod q.
\end{equation}
Then
\begin{equation}
C_2 - C_1 = M(T_2-T_1) \bmod q.
\end{equation}
Suppose that the first plaintext component is a nonce block and that the two plaintexts differ only by incrementing the nonce by one, while the remaining plaintext blocks remain unchanged. Then, except with negligible probability due to carry across the full encoded nonce, the plaintext difference is a basis vector, and the ciphertext difference reveals one entire column of the effective matrix.

For the target protocol, the useful values are $N_t$, $N_t+1$, $N_r$, and $N_r+1$. If, in a given session, these values are encrypted using the same effective matrix $M$ and the same modulus $q$, then the pairs $(N_t,N_t+1)$ and $(N_r,N_r+1)$ each reveal the same column of $M$. The second pair serves as a built-in consistency check: a wrong identification of the effective matrix or the modulus causes the two recovered columns to disagree with overwhelming probability.

\subsection{Why Reuse is Plausible}
In a conventional cryptographic design, reuse of the same effective transform would be vanishingly unlikely. Here the situation is different. In the lightweight setting, there are only eight AM choices, only four SUEO choices, and only a limited DBLTKM-generated matrix family. As a result, repeated effective matrices are not negligible events. Moreover, because DBLTKM explicitly includes transposition, a session may expose a column of $A$ while another session exposes a column of $A^T$, which provides additional information about the same underlying secret matrix.

\subsection{From Effective Matrices to the Secret Matrices}
The attack proceeds as follows.
\begin{enumerate}
    \item Collect many sessions and identify those sessions in which the ciphertext pairs corresponding to $(N_t,N_t+1)$ and $(N_r,N_r+1)$ are consistent with the same effective matrix and the same modulus.
    \item For each useful session, recover the leaked column of the corresponding effective matrix.
    \item Use the DBLTKM transpose option to relate observations obtained from $A$ and $A^T$, and similarly from $B$ and $B^T$.
    \item Combine these leaked columns across multiple sessions to recover three entries of each underlying $2\times 2$ secret matrix.
    \item Recover the remaining entry of each secret matrix from an ordinary ciphertext equation once the corresponding plaintext block and session modulus are known.
\end{enumerate}

The key point is that the attacker does not need to classify every possible DBLTKM construction explicitly. It is sufficient to exploit those sessions in which the effective matrix is reused and the nonce-difference equations are directly applicable. In the concrete $2\times 2$ setting, each useful session reveals one full column of the corresponding effective matrix. A leaked column of $A$ together with a leaked column of $A^T$ reveals three entries of the matrix $A$, and similarly for $B$.

To see how the remaining entry is recovered, write
\begin{equation}
A=\begin{bmatrix}a&b\\ c&d\end{bmatrix}.
\end{equation}
Suppose that the attack has already recovered $a$, $b$, and $c$. Consider any ordinary ciphertext equation of the form
\begin{equation}
\begin{bmatrix}C_1\\ C_2\end{bmatrix}=A\begin{bmatrix}X\\ Y\end{bmatrix}\bmod q.
\end{equation}
Then the second row gives
\begin{equation}
dY \equiv C_2-cX \pmod q.
\end{equation}
Whenever $Y$ is invertible modulo $q$, the missing entry is obtained as
\begin{equation}
d \equiv (C_2-cX)Y^{-1} \pmod q.
\end{equation}
Thus the nonce-difference phase only needs to recover three entries of each base matrix; the final entry follows in the subsequent plaintext/modulus recovery phase. Repeating this argument for both base matrices allows the attacker to reconstruct the full secret matrix family used by the protocol.

\begin{remark}
The matrix-recovery phase does not require recovering the full session modulus first. It suffices that the same effective matrix and the same modulus are used inside the session from which the column is recovered. This is the key reason we separate matrix recovery from modulus recovery.
\end{remark}

\section{Recovery of Candidate Session Moduli}\label{sec:modulus}
After the secret matrices have been recovered, the remaining task is to identify the candidate 64-bit moduli used by the AM mechanism. Our approach is to use multi-session consistency tests to narrow the set of plausible modulus candidates and then verify them against the full protocol equations.  The direct approach is to guess a 64-bit modulus, then test the its correctness through decryption, but the complexity of the direct approach is high. We will reduce this complexity significantly by considering a much smaller factor of $p$.  

\subsection{Motivation for a Factor-First Search}
In the target design, the AM mechanism is intended to provide eight distinct 64-bit candidate moduli derived from a common hidden quantity. In such a setting, it is natural to search first for smaller factors or residues that can be tested efficiently across many sessions, and only later reconstruct full 64-bit candidates.

For a 128-bit $p$ to have eight distinct 64-bit moduli, a small factor must appear in some 64-bit modulus $q$. Suppose that the smallest factor that appear in the 64-bit moduli is 64-bit, then there could be only two distinct 64-bit moduli.  If the smallest factor that appear in the 64-bit moduli is 32-bit, then $q$ has at most four distinct 32-bit factors, and $q$ have at most six distinct 64-bit factors.  Thus some factors less than 32-bit must appear in some 64-bit moduli.   
%This strategy is especially attractive in the lightweight regime studied in the protocol paper. The AM table is small, so the number of distinct moduli that can actually appear in practice is also small. Consequently, even partial information about a modulus can be useful, because it sharply reduces the set of candidates that must later be checked against the full ciphertext equations.

\subsection{Identification of the Reduced Modulus Value $q_x$}

At this stage, we do not assume that the effective matrix is symmetric. Let
\[
A=
\begin{pmatrix}
m_{11} & m_{12}\\
m_{21} & m_{22}
\end{pmatrix},
\qquad
A^T=
\begin{pmatrix}
m_{11} & m_{21}\\
m_{12} & m_{22}
\end{pmatrix}.
\]
Let the long-term secret be
\[
S=
\begin{pmatrix}
s_1\\
s_2
\end{pmatrix}.
\]
The purpose of this phase is to identify a correct reduced modulus value $q_x$ and then recover
\[
S \bmod q_x.
\]

The key observation is that the same long-term secret vector $S$ is reused across different sessions, while the protocol may apply either $A$ or $A^T$ in different sessions. Suppose that in one session the protocol uses $A$ to encrypt $S$, producing ciphertext
\[
C^{(A)}=
\begin{pmatrix}
c^{(A)}_1\\
c^{(A)}_2
\end{pmatrix}
\equiv
A
\begin{pmatrix}
s_1\\
s_2
\end{pmatrix}
\pmod{q_x}.
\]
This gives the reduced equations
\begin{align}
m_{11}s_1+m_{12}s_2 &\equiv c^{(A)}_1 \pmod{q_x}, \label{eq:Aqx1}\\
m_{21}s_1+m_{22}s_2 &\equiv c^{(A)}_2 \pmod{q_x}. \label{eq:Aqx2}
\end{align}

Now suppose that in another session the protocol uses $A^T$ to encrypt the same long-term secret vector $S$, producing ciphertext
\[
C^{(A^T)}=
\begin{pmatrix}
c^{(A^T)}_1\\
c^{(A^T)}_2
\end{pmatrix}
\equiv
A^T
\begin{pmatrix}
s_1\\
s_2
\end{pmatrix}
\pmod{q_x}.
\]
This gives
\begin{align}
m_{11}s_1+m_{21}s_2 &\equiv c^{(A^T)}_1 \pmod{q_x}, \label{eq:ATqx1}\\
m_{12}s_1+m_{22}s_2 &\equiv c^{(A^T)}_2 \pmod{q_x}. \label{eq:ATqx2}
\end{align}

Since the same secret vector $S$ is used in both sessions, these equations can be combined. For example, subtracting \eqref{eq:Aqx1} from \eqref{eq:ATqx1} gives
\[
(m_{21}-m_{12})s_2 \equiv c^{(A^T)}_1-c^{(A)}_1 \pmod{q_x}.
\]
Hence, whenever $m_{21}-m_{12}$ is invertible modulo $q_x$, the value of $s_2 \bmod q_x$ is uniquely determined. The value of $s_1 \bmod q_x$ then follows from any independent reduced equation above. More generally, the four reduced equations \eqref{eq:Aqx1}--\eqref{eq:ATqx2} form an overdetermined system in the two unknowns $s_1$ and $s_2$, and a correct candidate value of $q_x$ should make this system mutually consistent.

This gives a practical consistency test for candidate values of $q_x$. For each guessed $q_x$, the attacker searches among the collected sessions for one session in which $A$ is used to encrypt $S$ and another in which $A^T$ is used to encrypt the same long-term secret vector $S$. The attacker then reduces the corresponding ciphertext equations modulo $q_x$ and solves for $S \bmod q_x$. A correct guess of $q_x$ should produce mutually consistent reduced values of $S$ across multiple such session pairs, whereas an incorrect guess is generally not expected to do so.

Once the correct reduced modulus value $q_x$ has been identified, the attacker proceeds to reconstruct the corresponding candidate 64-bit modulus $q$. After $q$ is known, the remaining unknown entry of each base matrix can be recovered from the full ciphertext equations, and the remaining protocol secrets then follow by straightforward decryption.

\subsection{Reconstructing and Verifying 64-bit Modulus Candidates}
After several plausible reduced candidates have been identified, they are combined to form candidate 64-bit moduli. Because the target protocol uses only a small AM table, the number of such candidates remains manageable. The attacker can therefore enumerate the resulting 64-bit candidates and test them directly.

A candidate modulus $q$ is accepted only if it passes a stronger verification step: when the full session equations are solved under $q$, the resulting recoveries of the long-term secret $S$ and the associated plaintext blocks must be mutually consistent across many sessions. In other words, the reduced-equation phase serves only to filter candidates, whereas the final decision is made using the full protocol relations.

Once the correct 64-bit modulus $q$ has been identified, the recovery of the full matrices $A$ and $B$ is immediate. The nonce-difference and transpose analysis already determines three entries of each $2\times 2$ base matrix. The remaining fourth entry of each matrix is then obtained from an ordinary ciphertext equation, because both the effective matrix form and the modulus are now known. Therefore, at the end of this phase, both $A$ and $B$ are fully recovered, after which the remaining protocol secrets follow by straightforward decryption.

%This two-stage strategy is important. It avoids overstating what can be concluded from reduced equations alone, while still showing that the small AM state space makes modulus recovery realistic in the lightweight parameter regime. Once a candidate modulus passes the full consistency checks, it can be used in the next stage to recover the remaining protocol secrets.

\section{Full Recovery of the Remaining Secrets}\label{sec:fullbreak}
Once the full base matrices and the correct 64-bit modulus have been recovered, the remaining protocol secrets follow by straightforward decryption.

\subsection{Recovery of Long-Term and Session Secrets}
With the correct matrix and modulus available, the attacker decrypts the protocol messages and recovers
\begin{equation}
S,\quad ID,\quad N_t,\quad N_r.
\end{equation}
The update values $S^d,S^p,S^c$ are then recovered from the later encrypted messages. Strictly speaking, a complete impersonation break requires only the reduced indices
\begin{equation}
S^d \bmod 20,\qquad S^p \bmod 4,\qquad S^c \bmod 8,
\end{equation}
because these values fully determine the next DBLTKM pattern, SUEO order, and AM modulus in the lightweight setting. However, after matrix and modulus recovery, the full encoded values can also be obtained.

\subsection{Complete Protocol Compromise}
At this point the attacker can compute the same next-state updates as the legitimate parties, decrypt future traffic, and forge valid messages. Hence the protocol is fully broken in the standard operational sense: confidentiality of the protocol state is lost, mutual authentication can be bypassed, and future sessions can be predicted or impersonated.

\subsection{Complexity of the Attack}

The practicality of the attack relies on the small state space of the target design. The cost is dominated by collecting enough sessions to observe repeated effective matrices and by the search over candidate reduced modulus values. This search is pruned by a multi-session consistency test, which discards candidate reduced modulus values that do not yield mutually consistent reduced solutions for $S$ across sessions. Since the AM table contains only a small number of candidate 64-bit moduli and SUEO has only four orderings in the lightweight regime, the overall search remains manageable.

In Section IV, the probability that the same effective matrix $A$ and the same modulus $q$ are used to encrypt $N_t$, $N_r$, $N_t+1$, and $N_r+1$ in a given session is
\[
\frac{1}{4}\times\frac{1}{8}=\frac{1}{32}.
\]
The same probability applies to $A^T$, and likewise to $B$ and $B^T$. Therefore, any one target configuration of this form is expected to appear once every roughly 32 sessions on average. Consequently, the attacker must collect a moderate number of sessions until all required matrix/modulus configurations have appeared. The computational work in this phase is negligible compared with the session collection cost.

In Section V, the main computational cost is the search over candidate values of $q_x$. Under the structural assumption that one relevant factor of a 64-bit modulus has size at most 32 bits, the attacker tests at most about $2^{32}$ candidate values. For each candidate $q_x$, the attacker checks consistency across a small number of useful session instances by solving small modular linear systems for $S \bmod q_x$. Since a useful configuration appears with probability about $1/32$, only a moderate number of such systems need to be tested per guess; for example, testing around 128 small systems per candidate gives a total computational effort on the order of
\[
2^{32}\cdot 128 = 2^{39}
\]
small modular linear-system checks and the correct guess is expected the correct guess is expected to yield about four mutually consistent reduced solutions on average. 

In Section VI, once the correct modulus and the required matrix entries have been identified, the remaining recovery steps involve only straightforward solution of a few small linear equations and direct decryption of the remaining protocol values. Hence the computational cost of this phase is negligible compared with the session-collection cost of Section IV and the candidate search in Section V.

Overall, the attack requires collecting a moderate number of sessions until the necessary matrix/modulus configurations appear, and then performing about $2^{39}$ small modular consistency checks in the reduced-modulus search phase. Thus the dominant computational complexity is the Section V search, while the dominant data complexity is the session collection in Section IV.

\section{Discussion}
Our attack highlights a general lesson that extends beyond the target protocol. Replacing standard cryptographic primitives with highly structured linear transformations often produces designs that look combinatorially rich but are in fact extremely small once implementation constraints are taken into account. In the present case, the requirement of lightweight deployment forces the number of secret matrices to remain tiny. As a result, the advertised update mechanisms do not provide a cryptographically meaningful state space.

A second lesson is that correctness constraints matter. If a protocol claims to encrypt and decrypt 128-bit values modulo a 64-bit modulus, then either a special encoding must be specified or exact decryption is impossible. Any cryptanalysis of such a design must therefore consider not only the nominal bitlengths claimed in the protocol description but also the actual coordinate ranges required for correctness.

Finally, the target protocol illustrates the danger of using transposition and block-diagonal composition as a substitute for real diffusion. These operations do not hide the linear structure; instead, they create algebraic relations that can be exploited across sessions.

\section{Conclusion}\label{sec:conclusion}
We have presented a cryptanalysis of a recently proposed lightweight RFID authentication protocol based on AM, SUEO, and DBLTKM matrix updates. Our analysis shows that the protocol has a very small effective state space in the concrete lightweight setting, suffers from a correctness inconsistency in its treatment of 128-bit values under 64-bit modular encryption, and is vulnerable to a practical multi-session algebraic attack. By leveraging repeated use of the same effective matrix and modulus, we recover columns of the effective matrices from the encryptions of $N_t$, $N_t+1$, $N_r$, and $N_r+1$. Using transpose relations across sessions, we recover three entries of each underlying secret matrix, and then determine the remaining entries from ordinary ciphertext equations once the corresponding plaintext blocks and moduli are known. We then use multi-session consistency tests on reduced equations for the long-term secret $S$ to identify candidate session moduli, which are subsequently verified against the full protocol equations and used to recover the remaining protocol secrets.

These results indicate that the security claims of the target protocol do not hold in the lightweight parameter regime advocated by its authors. More broadly, our work emphasizes that lightweight authentication protocols must be evaluated not merely by storage savings or combinational key-space counts, but by rigorous cryptanalysis of their actual algebraic structure.

\bibliographystyle{IEEEtran}

\end{document}